\documentclass[twocolumn,pre,superscriptaddress,floatfix,longbibliography]{revtex4-1}

\usepackage{graphics}
\usepackage{graphicx}
\usepackage{amsmath}
\usepackage{amssymb,xcolor}
\usepackage{gensymb}
\usepackage{ulem}

\newcommand{\textin}[1]{\mbox{\scriptsize{#1}}}

\definecolor{grisclair}{rgb}{0.6,0.6,0.6}

\newcommand{\beq}{\begin{equation}}
\newcommand{\ee}{\end{equation}}

\begin{document}

\title{Critical bubble bursting in real water. Effect of surface-active contaminants}
\author{S. Rodríguez-Aparicio}
\address{Depto.\ de Ingenier\'{\i}a Mec\'anica, Energ\'etica y de los Materiales and\\ 
Instituto de Computaci\'on Cient\'{\i}fica Avanzada (ICCAEx),\\
Universidad de Extremadura, E-06006 Badajoz, Spain}
\author{A. Cebrián-García}
\address{Depto.\ de Ingenier\'{\i}a Mec\'anica, Energ\'etica y de los Materiales and\\ 
Instituto de Computaci\'on Cient\'{\i}fica Avanzada (ICCAEx),\\
Universidad de Extremadura, E-06006 Badajoz, Spain}
\author{E. J. Vega}
\address{Depto.\ de Ingenier\'{\i}a Mec\'anica, Energ\'etica y de los Materiales and\\ 
Instituto de Computaci\'on Cient\'{\i}fica Avanzada (ICCAEx),\\
Universidad de Extremadura, E-06006 Badajoz, Spain}
\author{J. M. Montanero}
\address{Depto.\ de Ingenier\'{\i}a Mec\'anica, Energ\'etica y de los Materiales and\\ 
Instituto de Computaci\'on Cient\'{\i}fica Avanzada (ICCAEx),\\
Universidad de Extremadura, E-06006 Badajoz, Spain}
\author{M. G. Cabezas}
\address{Depto.\ de Ingenier\'{\i}a Mec\'anica, Energ\'etica y de los Materiales and\\ 
Instituto de Computaci\'on Cient\'{\i}fica Avanzada (ICCAEx),\\
Universidad de Extremadura, E-06006 Badajoz, Spain}

\begin{abstract}
We study the bursting of a bubble on a liquid free surface under critical conditions, i.e., those leading to the minimum (maximum) size (velocity) of the first-emitted jet droplet. Our experiments show that a tiny amount of surfactant considerably increases (decreases) the droplet radius (velocity). The volume of the first-emitted droplet increases by a factor of 20 for a concentration that produces an insignificant reduction in the bubble surface tension. The total liquid volume ejected by the bubble increases with the surfactant concentration. Surfactant accumulates at the bubble base due to cavity bottom shrinkage and surfactant convection. The resulting reduction in surface tension narrows the region of free surface reversal. Despite this effect, the size of the emitted droplet increases due to the Marangoni stress acting on the jet surface. Marangoni stress slows down the interface of the liquid jet, delaying the detachment of the droplet. More liquid flows into the droplet, increasing the mass and energy transfer to the resulting spray. A significant increase in the droplet size is also observed with a weak surfactant. This indicates that natural water contamination can substantially alter the bursting of bubbles under critical conditions. Our results may explain the size of the particles emitted by bubble bursting in seawater.
\end{abstract}

\maketitle



\section{Introduction}
\label{Intro}

When an air bubble rises to the water surface due to buoyancy, the cap film covering the bubble breaks up. The following collapse of the bubble cavity produces capillary waves that converge at the cavity base. This creates a Worthington jet whose tip emits tiny liquid drops \citep{DPJZ02,GS16,G17a,BBWFYB18,DGLDZPS18,GR19,GL21}. The jet drops coming from the bursting of tiny bubbles on the ocean surfaces are a widely accepted origin of aerosols around 0.1 $\mu$m in size \citep{V15b,D22, WDMR17}. These aerosols act as cloud condensation nuclei in atmospheric regions where water vapour concentration reaches supersaturation \citep{Cochran2017,Meshkhidze2019}. The jet drops also transport chemicals (toxins/microplastics) \citep{Wetal17,SLND23} and biological substances (bacteria/viruses) \citep{B21} into the atmosphere, which has obvious consequences for public health. 

The jet droplets' small size and large vertical velocity are critical in the phenomena described above. The parameter conditions leading to the minimum size and maximum velocity correspond to $\text{Bo}=0$ and $\text{La}=\text{La}^*\simeq 1110$, where $\text{Bo}=\rho g R_b^2/\sigma$ is the Bond number, $\text{La}=\rho R_b \sigma/\mu^2$ is the Laplace number, $\rho$, $\mu$, and $\sigma$ are the liquid density, viscosity and surface tension, $R_b=[3V/(4\pi)]^{1/3}$ and $V$ are the bubble radius and volume, and $g$ is the gravity. The existence of an optimum value of the Laplace number, La$^*$, has been attributed to an energy-focusing effect \citep{DGLDZPS18} during the bubble collapse: the mechanical energy per unit volume focused on the ejected ligament is maximised for that optimum value of the Laplace number \citep{GL21}. 

Bubble bursting in a clean liquid \citep{DPJZ02,GS16,G17a,BBWFYB18,DGLDZPS18,GR19,GL21} is an idealisation of the more complex phenomenon taking place in real fluids, such as seawater. Contaminants in the marine boundary layer, both natural and anthropogenic, significantly impact the bubble bursting process and, consequently, the emission of jet droplets \citep{Hardy1982,Masry2021}. Surfactants \citep{Baryiames2021,ND21,NED21,CKBSCJM21,PPS22,JIWEF23,VM24,YBJF24}, polymeric molecules \citep{SLJ21,TSKGMFM21,JYWEF23,RRMGC23,YBJF24,DOZLS25,BSJVT24,CMLCVM25,BYF25}, oils \citep{JYF21,YJF23,YJAF23,YLF25} and particles of different nature \citep{JSF22,DMB23} have proved to substantially affect bubble bursting. 

Previous experiments of bubble bursting with surfactants have shown the noticeable effect of these substances on the size and velocity of the ejected droplet \citep{PPS22,VM24,YBJF24}. Those experiments were conducted within a parameter region far from the critical conditions $\text{Bo}=0$ and $\text{La}=\text{La}^*$. Under these conditions, strong surface convection of surfactant over the interface and fast compression of the bubble bottom may considerably increase the surfactant concentration in that critical region, enhancing the surfactant effects. In this work, we conduct experiments for $\text{Bo}\simeq 0$ and $\text{La}\simeq \text{La}^*$ with both moderately strong and weak surfactants. The latter simulates the unavoidable presence of surface-active contaminants in water. 



\section{Experimental procedure}
In our experiments, a needle was located at the bottom of a tank filled with the working liquid. The bubble detached from the needle, rose across the tank until reaching the free surface, and burst. The bubble bursting was recorded with a high-speed video camera. Experiments were repeated five times to ensure their reproducibility. For more details on the experimental setup, refer to the Supplemental Material.

Experiments involving critical bursting with water are challenging because of the tiny spatial and temporal scales characterising the phenomenon (bubbles must be approximately 20 $\mu$m in radius and emit droplets 400 nm in radius at speeds larger than 115 m/s). Instead, we work with dimethyl sulfoxide (DMSO)/water mixtures and bubble radii in the interval $200-700$ $\mu$m to produce bubble bursting with $\text{Bo}\lesssim 0.1$ and a range of La including $\text{La}^*$. It is worth mentioning that our experiments with glycerol/water mixtures (commonly used in this kind of experiment) were not sufficiently reproducible and resulted in larger jet droplets (see the Supplemental Material).

\begin{figure*}
\centering
\includegraphics[height=4.35cm]{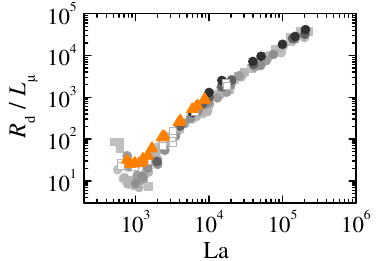} 
\includegraphics[height=4.35cm]{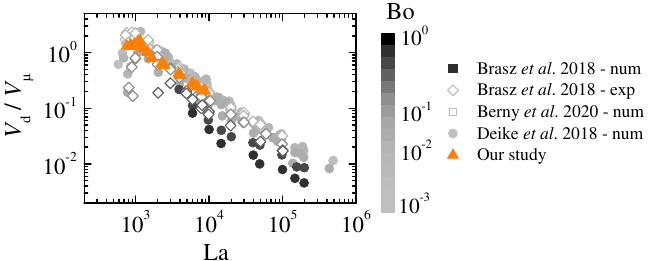}
\caption{Radius $R_d$ and velocity $V_d$ of the first emitted droplet as a function of the Laplace number for surfactant-free experiments. The droplet radius and velocity are measured in terms of the viscous-capillary length $L_{\mu}=\mu^2/(\rho\sigma)$ and velocity $V_{\mu}=\sigma/\mu$. The triangles are our experimental data for $\text{Bo}=0.007-0.018$. The squares, circles, and diamonds are the results of \citet{BBWFYB18}, \citet{BDSP20}, and \citet{DGLDZPS18}, respectively.}
\label{CleanRes}
\end{figure*}

To validate our experimental procedure, we performed experiments with a clean interface. Figure \ref{CleanRes} shows the radius $R_d$ and velocity $V_d$ of the first-emitted droplet as a function of the Laplace number for DMSO/water mixtures in the absence of surfactant. Our results are in excellent agreement with previous numerical simulations \citep{BBWFYB18,BDSP20}. A sharp minimum in $R_d/R_b$ was found for $\text{La}\simeq 1100$. The minimum value of the first-emitted droplet radius corresponds to $R_d/R_b\simeq 0.023$, similar to those found in previous carefully conducted experiments \citep{BBWFYB18}. This result agrees with the prediction $R_d/R_b\simeq 0.0239$ for $\text{La}=1275$ obtained from the scaling law derived by \citet{GL21}. In some experiments, the size of droplets subsequently emitted was even smaller than that value (see the Supplemental Material). As mentioned above, critical bubble bursting in water occurs with bubbles around 20 $\mu$m in radius. This means that bubble bursting in water can produce droplets with radii well below 500 nm.

\begin{figure*}
\centering
\includegraphics[width=0.24\textwidth]{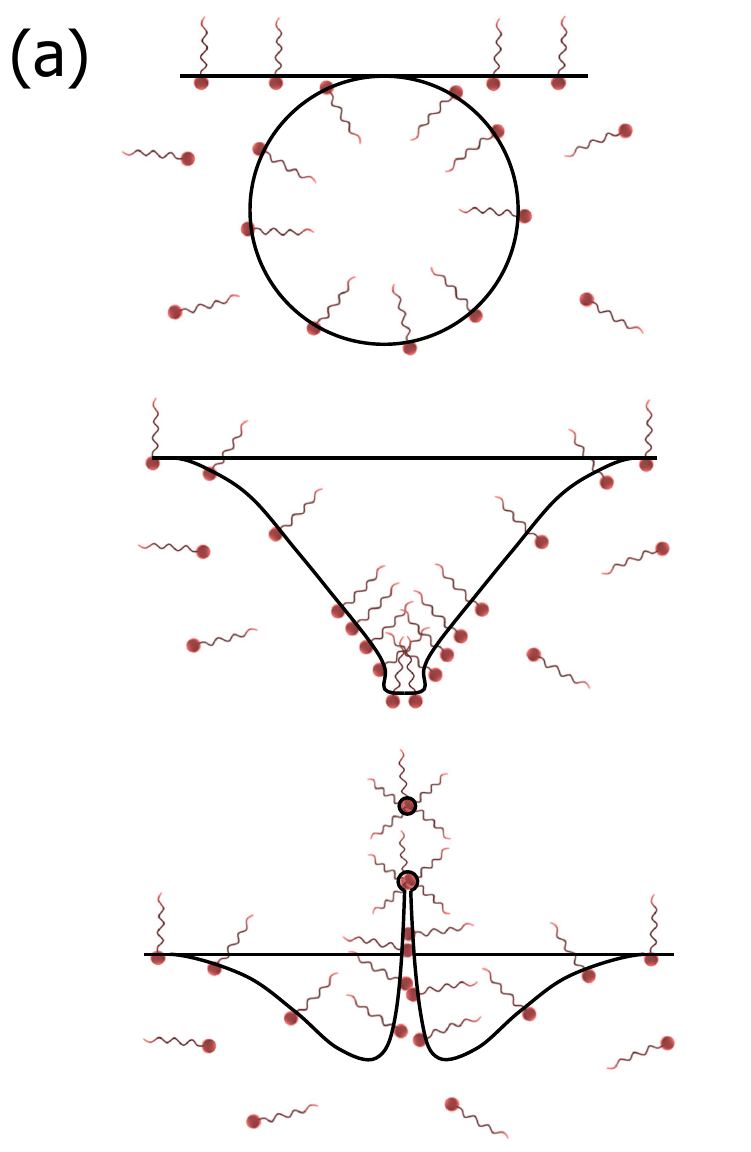}\includegraphics[height=4cm]{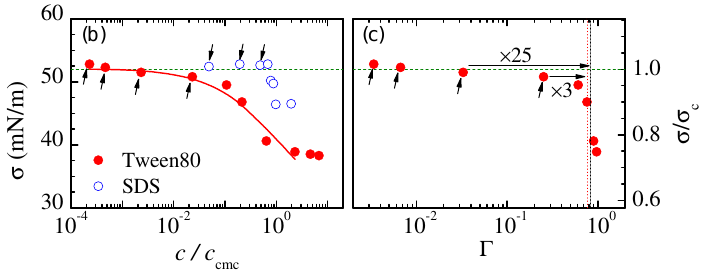}
\caption{(a) Bubble bursting in the presence of surfactant. (b) Surface tension $\sigma$ as a function of the surfactant concentration $c/c_{\textin{cmc}}$ for DMSO/water mixtures (50/50 wt). The arrows indicate the surfactant concentrations considered in this study. The solid line is a fit to the experimental data using the Langmuir equation of state for Tween 80 (see the Supplemental Material). (c) Surface tension $\sigma$ as a function of the surfactant coverage $\Gamma$ calculated from the fit for Tween 80. The horizontal arrows indicate the estimated increase in the surface coverage during the bubble bursting, as explained in the text.}
\label{Fig1New}
\end{figure*}

Suppose we neglect the effect of the gas. In that case, the bubble bursting in the absence of surfactant is characterized
by the Bond and Laplace numbers. In our experiments, the bubble remains at rest on the free surface for significantly longer than the times required for surfactant adsorption and desorption. This means that the surface density $\widehat{\Gamma}$ of both the bath and bubble surfaces takes the equilibrium value $\widehat{\Gamma}_{\textin{eq}}$ before the bubble bursting. Then, the liquid film between the bubble and the bath surface drains and breaks up, and the bubble bursting begins (Fig.\ \ref{Fig1New}a). The entire process leading to the formation of the jet droplets takes a time much shorter than the characteristic adsorption and desorption times \citep{RABK09,MK12,KWCCAB20}. This implies that the surfactant adsorbed onto the interface at the initial instant remains in the monolayer during the bubble bursting. In this sense, the surfactant can be regarded as insoluble even though it is present as a solute in the liquid phase. 

The surface diffusion time $t_D=R_b^2/D_s$ ($D_s$ is the surface diffusion coefficient) is much larger than the characteristic time of the process. Therefore, surfactant diffusion over the interface can be ignored. Under the above conditions, the dimensionless number characterizing the surfactant effect are typically the equilibrium value of the surface coverage $\Gamma_{\textin{eq}}=\widehat{\Gamma}_{\textin{eq}}/\widehat{\Gamma}_{\infty}$ ($\widehat{\Gamma}_{\infty}$ is the maximum packing fraction) and the Marangoni number $\text{Ma}=\widehat{\Gamma}_{\infty}R_g T/\sigma_c$ ($\sigma_c$ is the clean interface surface tension). The equilibrium isotherm $\widehat{\Gamma}(c)$ ($c$ is the surfactant volumetric concentration) is unknown for the mixtures considered in our analysis, which prevents us from determining $\Gamma_{\textin{eq}}$ and Ma. Alternatively, we use $c/c_{\textin{cmc}}$ ($c_{\textin{cmc}}$ is the critical micele concentration) and $\beta=(\sigma_c-\sigma(c_{\textin{cmc}}))/\sigma_c$ to characterize the surfactant monolayer.

To analyse the effect of the surfactant, we selected two surfactants with significantly different strengths: $\beta=0.11$ for sodium dodecyl sulfate (SDS) and $\beta=0.26$ for Tween 80 (Fig.\ \ref{Fig1New}b). Given its small strength, SDS can represent the presence of impurities in water. Tween 80 in the DMSO/water mixtures considered here can be regarded as a moderately weak surfactant (e.g., $\beta=0.47$ for SDS in water). The equilibrium surfactant concentrations considered in our analysis were so small that the surface tension was practically the same as that of the clean interface (Fig.\ \ref{Fig1New}b). In the Tween 80 case, $\sigma(c)$ could be fitted by the Langmuir equation of state (Fig.\ \ref{Fig1New}b). This allowed us to estimate the isotherm $\Gamma(c)$ and the equation of state $\sigma(\Gamma)$ (Fig.\ \ref{Fig1New}c) (see the Supplemental Material for details). For this surfactant, $\Gamma_{\textin{eq}}\leq 0.26$ in our bubble bursting experiments. 

\section{Results}
\label{Results}

\subsection{Cavity collapse}

\begin{figure*}
\begin{center}
\resizebox{\textwidth}{!}{\includegraphics{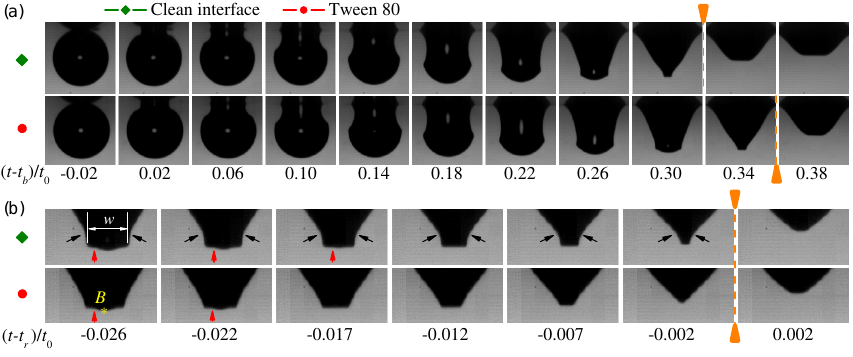}}
\resizebox{\textwidth}{!}{\includegraphics{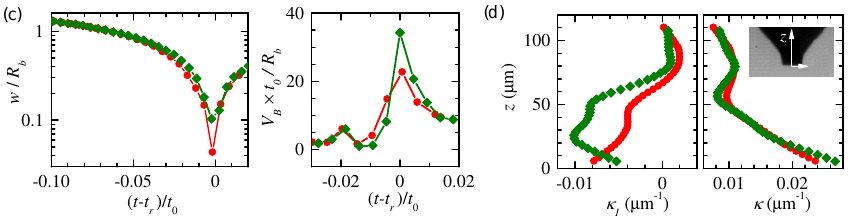}}
\end{center}
\caption{(a) Images of the cavity collapse without surfactant and with Tween 80 at $c/c_{\textin{cmc}}=0.0233$ ($\Gamma_{\textin{eq}}=0.26$) both for $\text{La}\simeq \text{La}^*$ and $\text{Bo}\simeq 0.01$. (b) Zoom in on the bubble bottom region close to the free surface reversal. The labels indicate the time to the film rupture (a) and free surface reversal (b) divided by the inertio-capillary time $t_0$. The orange arrows indicate the free surface reversal instant. The red arrows in (b) point at a previous capillary wave. The black arrow in (b) indicates the free surface curvature $\kappa_1=-d^2r/dz^2/[1+(dr/dz)^2]^{3/2}$ partially eliminated by the surfactant. (c) Cavity bottom width $w$ and upward velocity $V_B$ as a function of time to the free surface reversal. (d) Curvature $\kappa_1$ and total curvature $\kappa=\kappa_1+\kappa_2$ ($\kappa_2=[r\sqrt{1+(dr/dz)^2}]^{-1}$) along the lateral free surface (excluding the corner and bottom of the cavity) for $(t-t_r)/t_0=-0.07$ calculated with a subpixel resolution technique \citep{M24}. The bubble was  225 $\mu$m in radius.}
\label{images}
\end{figure*}

As shown below, the surfactant has a negligible effect on the critical Laplace number $\text{La}^*$. In this section, we analyze the impact on the cavity evolution of Tween 80 at $c/c_{\textin{cmc}}=0.0233$ ($\Gamma_{\textin{eq}}=0.26$) for $\text{La}=\text{La}^*$. Due to the small surfactant concentration, the time elapsed between film rupture and the free surface reversal increased only around 12\% in the presence of Tween 80 (Fig.\ \ref{images}a). The cavity shape remained largely unchanged by the surfactant until $(t-t_b)/t_0\simeq 0.2$ ($t_b$ is the instant of the film rupture and $t_0=(\rho R_b^3/\sigma_c)^{1/2}\simeq 0.50$ ms is the inertio-capillary time). Significant differences appeared for $(t-t_b)/t_0\gtrsim 0.2$.

Figure \ref{images}b zooms in on the cavity bottom during the last stage of the cavity collapse. The instant $t=t_r$ of the free surface reversal is determined as that for which the velocity of the cavity bottom is maximum (Fig.\ \ref{images}c). In the absence of surfactant, the width $w$ of the cavity bottom takes the value $w/R_b\simeq 0.1$ at $(t-t_r)/t_0\simeq -0.002$, which is consistent with previous simulation results \citep{GL21}. This value is smaller than the minimum one reported by \citet{EAFPSD25} for a larger Laplace number $\text{La}=2400$ without surfactant. This discrepancy may indicate a strong effect of viscosity on the free surface reversal at the critical conditions.

During the cavity collapse, the surfactant is convected toward the cavity bottom. The cavity bottom sharply shrinks before the interface reversal. Both effects contribute to increasing the surface concentration of surfactant in that region \citep{CKBSCJM21,PKSCJM24}. This produces a local reduction in surface tension, which narrows the width of the cavity bottom (Fig.\ \ref{images}c). The Marangoni stress opposes the flow converging to the cavity bottom. Both local solute-capillarity and Marangoni stress delay the free surface reversal (Fig.\ \ref{images}a).

The surface tension variation is very low for $\Gamma<0.6$ (Fig.\ \ref{Fig1New}c). Therefore, solute-capillarity and Marangoni stress are expected to be significant only when the local surfactant concentration rises above that value. This means that the local concentration eventually exceeds $2.5\Gamma_{\text{eq}}$ for $c/c_\text{cmc}=0.023$ ($\Gamma_\text{eq}\simeq 0.26$). Assume that there is a phase of the cavity collapse in which the concentration at the bubble bottom lies in the interval $2.5\Gamma_{\text{eq}} \lesssim \Gamma \lesssim 3\Gamma_{\text{eq}}$. In this phase, the surface tension reduction is less than 10\% (Fig.\ \ref{Fig1New}c). However, $\sigma(\Gamma)$ exhibits a relatively sharp dependence on $\Gamma$ (Fig.\ \ref{Fig1New}c). Therefore, Marangoni stress must be the major effect at this stage. This is consistent with the numerical simulations of \citet{CKBSCJM21} for a similar initial coverage and $\text{La}=2\times10^4$. In these simulations, the concentration at the bubble bottom reaches almost three times the equilibrium value during the final stage of the cavity collapse (see Fig.\ 2 in \citep{CKBSCJM21}). In our experiment, the phase influenced only by the Marangoni stress is followed by a last stage before the free surface reversal, in which the concentration increases further. In this last stage, local solute-capillarity contributes to narrowing the cavity bottom (Fig.\ \ref{images}c) (the Supplemental Material shows the high reproducibility of this result). Our results qualitatively agree with the numerical simulations of \citet{EAFPSD25}, which show that an accumulation of surfactants near the corner of the cavity results in a Marangoni stress that smooths the corner (Fig.\ \ref{Fig1New}b).

Figure \ref{images}d shows that the surfactant partially eliminates the interface curvature $\kappa_1=-d^2r/dz^2/[1+(dr/dz)^2]^{3/2}$ along one of the principal radii of curvature of the lateral bubble surface (see the black arrows in Fig. \ref{images}b). This increases the free surface radius $r(z)$ and therefore decreases the curvature $\kappa_2=[r\sqrt{1+(dr/dz)^2}]^{-1}$ along the other principal radius. The combination of these effects increases the bubble curvature $\kappa=\kappa_1+\kappa_2$ (decreases the capillary pressure $p_c=-\sigma\kappa$) next to the bottom. This effect was also observed in simulations for larger values of La \citep{PKSCJM24}. The curvature $\kappa_1$ may be affected by numerical error due to the limited spatial resolution of the image, even though the free surface location was determined at the subpixel level \citep{M24}.


\subsection{Droplet emission}

\begin{figure*}
\begin{center}
\includegraphics[height=4.5cm]{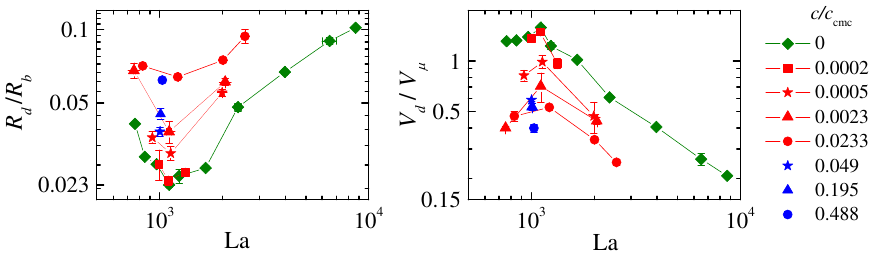}
\end{center}
\caption{Dimensionless radius $R_d/R_b$ and velocity $V_d/V_\mu$ ($V_\mu=\sigma_c/\mu$) of the first-emitted jet droplet as a function of the Laplace number La without surfactant (green symbols), with Tween 80 (red symbols), and with SDS (blue symbols).  The error bars indicate the standard deviation.}
\label{merit}
\end{figure*}

We now examine the effect of the surfactant on spray production. Figure \ref{merit} shows the radius $R_d$ and velocity $V_d$ of the first-emitted droplet as a function of the Laplace number for a clean interface and different concentrations of Tween 80 and SDS. The surfactant hardly changes the critical Laplace number $\text{La}^*$ for which the droplet radius is minimum. For a fixed Laplace number, the droplet radius monotonically increases and the velocity monotonically decreases with the surfactant concentration $c/c_{\textin{cmc}}$ (Fig.\ \ref{merit}). The same effect was observed for $\text{La}=5.4\times 10^4$ and lower SDS concentrations than those for which ejection was suppressed \citep{VM24}.

The phenomenon observed here is substantially different from that occurring with saturated monolayers and large Laplace numbers. Experiments under those conditions have demonstrated that a surfactant dissolved at the appropriate concentration dampens the short-wavelength capillary waves arising during the cavity collapse \citep{RRMGC23,VM24}, thereby intensifying the focusing of energy and enhancing jet ejection. This results in a significant increase (decrease) in the first-emitted droplet velocity (radius) \citep{RRMGC23,VM24}. A tiny capillary wave can be observed at the base of the clean cavity (see the red arrows in Fig.\ \ref{images}a), indicating a slight deviation with respect to the critical condition $\text{La}=\text{La}^*$. The surfactant does not produce a significant damping effect. As a consequence, we do not observe the non-monotonous dependence of the droplet radius and velocity on the surfactant concentration described by \citet{RRMGC23} and \citet{VM24}.

Interestingly, there is a significant increase in the droplet radius even at concentrations where the equilibrium surface tension is practically the same as that of the clean interface. For instance, $R_d$ increases by a factor of 2.8 at the concentration $c/c_{\textin{cmc}}=0.0233$ of Tween 80, for which $\sigma/\sigma_c=0.98$. In this case, the bubble bottom width $w$ reaches a minimum value much smaller than that without surfactant (Fig.\ \ref{images}b-c). This indicates that there are other effects in the jet (occurring after the free surface reversal) that contribute to increasing the droplet radius. \citet{VM24} have experimentally shown that the surfactant molecules are convected toward the jet tip so that the density of the monolayer covering the first-emitted droplet increases significantly for $\text{La}>\text{La}^*$. Simulations for large Laplace numbers show that the resulting Marangoni stress generates a recirculation zone near the jet base, which counteracts the flow from the jet bottom toward the front, contributing to increasing the radius of the droplet \citep{CKBSCJM21,PKSCJM24}. This mechanism also explains the increase in the droplet radius in our experiments.

The droplet radius increases and the velocity decreases with the surfactant strength $\beta$. For $\text{La}\simeq \text{La}^*$, the increase in the droplet radius due to the surfactant, $(R_d/R_b)-(R_d/R_b)_0$ (here $(R_d/R_b)_0$ is the dimensionless radius without surfactant), can be fitted by the power law $(R_d/R_b)-(R_d/R_b)_0= 17.9\, \beta^{2.5}\, (c/c_{\textin{cmc}})^{0.65}$ (see the Supplemental Material), indicating the importance of the surfactant strength $\beta$. A significant increase in the droplet size is also observed with a weak surfactant. This means that impurities in water substantially increase the minimum radius of the jet droplets. Bubble bursting in pure water is an idealisation that significantly differs from that occurring in nature. Our experiments suggest that the presence of impurities in real seawater results in larger jet droplets. This is consistent with the results obtained by \citet{WDMR17} using natural seawater. In their experiments, bubbles with a radius in the range 20-40 $\mu$m produced a significant number of dried particles with radii down to $\approx 0.5$ $\mu$m. The radii of the jet droplets that originated those particles were around twice those of the particles; i.e., 2.2 times the minimum radius in clean water. 

The first-emitted droplet velocity significantly decreases with the addition of Tween 80. However, the momentum of the droplet at $c/c_{\textin{cmc}}=0.0233$ is around seven times the value without surfactant. Interestingly, the droplet kinetic energy increases by a factor of 2.5 when the surfactant is added. This may reflect the decrease in the interfacial energy.

\begin{figure*}
\begin{center}
\resizebox{.75\textwidth}{!}{\includegraphics{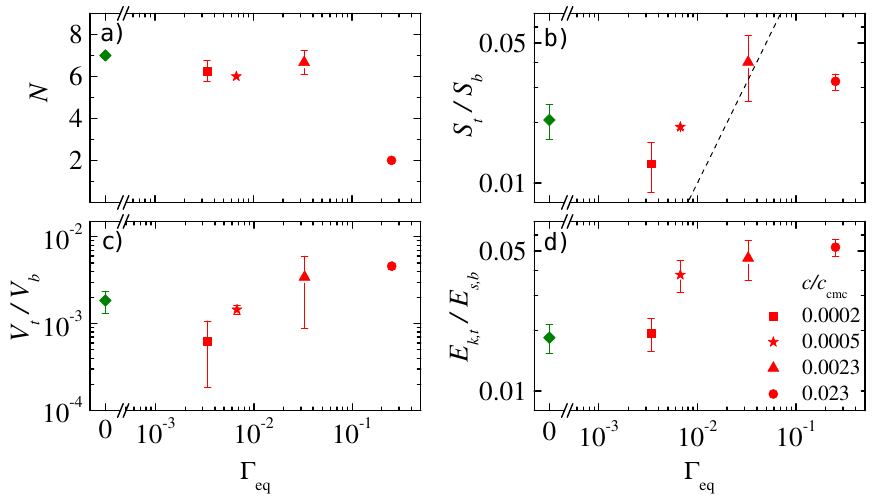}}
\end{center}
\caption{Number $N$ of jet droplets (a), total emitted surface $S_t$ (b), volume $V_t$ (c), and kinetic energy $E_{k,t}$ (d). The results are expressed in terms of the bubble surface $S_b=4\pi R_b^2$, volume $V_b=4/3\pi R_b^3$, and interfacial energy $E_{s,b}=\sigma_c S_b$. The dashed line in b) corresponds to $S_t/S_b=\Gamma_{\textin{eq}}$.}
\label{Alldrops}
\end{figure*}

According to \citet{BDSP20}, the bursting of each bubble can produce up to fourteen droplets that may contribute to spray formation. Figure \ref{Alldrops} shows the number $N$ of emitted droplets, their total surface $S_{t}$, volume $V_{t}$, and kinetic energy $E_{k,t}$ at the critical Laplace number for a clean interface and different concentrations of Tween 80. The combination of the Langmuir equation and the Gibbs isotherm for this surfactant allows us to calculate $\Gamma_{\text{eq}}(c)$, which allows us to plot the results versus this quantity. The results shown in Fig.\ \ref{Alldrops} correspond only to the droplets emitted with an upward velocity, excluding those produced by the fragmentation of the retracting jet. The radius and velocity of each droplet are shown in the Supplemental Material. The number of drops $N=7$ for a clean interface is in good agreement with numerical simulations for almost the same Bond and Laplace numbers ($\text{Bo}=0.015$ and $\text{La}=1142$) (see Fig.\ 5b in \citep{BDSP20}). It is worth noting that $N$ is very sensitive to the Bond and Laplace number for $\text{La}\simeq \text{La}^*$. In fact, the simulations of \citet{BDSP20} showed the ejection of fourteen droplets for $\text{La}=1430$ and $\text{Bo}=0.009$.

Our results demonstrate high reproducibility in the number of droplets, but significant scattering in the total emitted surface and volume due to the nonlinearity of the jet fragmentation process. The addition of surfactant reduces the number of emitted drops. Only two drops are emitted for the largest concentration $c/c_\text{cmc}=0.023$. This is consistent with the experimental results of \citet{VM24} for larger values of La, who showed the existence of a concentration interval without droplet emission.

The emitted surface $S_t$ and volume $V_t$ exhibits a minumum at $\Gamma_{\textin{eq}}=3.4 \times 10^{-3}$. Beyond this concentration, the surface, volume, and kinetic energy of the liquid released into the atmosphere increase with the surfactant concentration. The increase in kinetic energy cannot be attributed only to a decrease in the droplet surface tension. In fact, the kinetic energy of the surfactant-laden droplets exceeds the sum of the kinetic and interfacial energies in the clean interface case. Therefore, the surfactant enhances the transfer of energy to the spray. The increase in the mass and energy of the emitted droplets may explain the increase in the microplastic transference to the atmosphere caused by bubbles bursting in the presence of surfactants \citep{Masry2021}.

The surfactant initially adsorbed at the bubble's free surface accumulates at the bottom of the bubble during the collapse of the cavity. Let $S^*$ be the area of the surface that could be covered with surfactant at the maximum packing fraction. Mass conservation yields $S^*/S_b=\Gamma_\text{eq}$, where $S_b=4\pi R_b^2$ is the bubble surface area. For $c/c_\text{cmc}\leq0.0023$, $S^*$ is smaller than the total surface of the emitted droplets (Fig.\ \ref{Alldrops}b). For these concentrations, we can assume that practically all the surfactant adsorbed at the bubble surface is transported by the emitted droplets. Under this assumption, the mean surfactant coverage in the emitted droplets, $\langle\Gamma\rangle$, can be calculated from the conservation equation $\langle\Gamma\rangle S_t=\Gamma_\text{eq}S_b$. For $c/c_\text{cmc}=0.0023$ ($\Gamma_\text{eq}=0.033$), $\langle\Gamma\rangle$ becomes almost 25 times $\Gamma_\text{eq}$. The corresponding surface tension value, $\sigma (\langle\Gamma\rangle)$, is arbout 0.85$\sigma_c$ (Fig. \ref{Fig1New}c). This means that the reduction in surface tension in the jet is relatively small or affects only a small region. However, the values of $\langle\Gamma\rangle$ lie in an interval where $\sigma$ exhibits a significant dependence on $\Gamma$ (Fig. \ref{Fig1New}c), giving rise to significant Marangoni stress. We may conclude that the primary effect of the surfactant on jet dynamics is Marangoni stress, which slows the interface and delays droplet detachment. This allows for more liquid to flow inside the droplet, increasing the mass and energy transfer to the resulting spray.

\section{Conclusions}

In summary, we have demonstrated that a tiny amount of surfactant significantly increases the size and total mass of the emitted droplets through bubble bursting, even when its presence is not detected in surface tension measurements or does not significantly affect cavity collapse time. Surfactant accumulates at the bubble base due to the cavity bottom shrinkage and surfactant convection. Both solute-capillarity and  Marangoni stress alter the self-similar bubble collapse before the free surface reversal. This effect is also observed with a very weak surfactant, indicating that natural water contamination can substantially alter bubble bursting at the critical conditions. 

Our work provides evidence highlighting the significant influence of surface-active contaminants, commonly present in natural environments, on the singular dynamics of bursting bubbles —a critical phenomenon for the production of aerosols and the transportation of chemical and biological substances into the atmosphere.

\section*{Acknowledgement}

This work was financially supported by the Spanish Ministry of Science, Innovation and Universities (grant no. PID2022-140951OB-C22/AEI/10.13039/501100011033/FEDER, UE).

The authors report no conflict of interest.


%

\end{document}